\documentclass[12pt]{article}
\usepackage{graphicx}
\usepackage{epsf,amsmath,bbold,amsfonts,stmaryrd}

\usepackage[utf8]{inputenc}
\usepackage{mathrsfs}
\usepackage{appendix}
\usepackage{amssymb}
\usepackage{float}
\usepackage{color} 
\usepackage{cite}
\usepackage[colorlinks]{hyperref}
\hypersetup{pageanchor=false}
\usepackage{indentfirst}
\usepackage{url}
\usepackage{float}
\usepackage{caption}
\usepackage[numbers,square,comma,sort&compress,merge]{natbib}

\def\a{\alpha}

\def\d{\delta}
\def\D{\Delta}

\def\ep{\epsilon}

\def\f{\frac}

\def\G{\Gamma}

\def\l{\left}

\def\mc{\mathcal}

\def\m{\mu}
\def\n{\nu}

\def\nn{\nonumber}

\def\p{\partial}

\def\r{\right}
\def\s{\sigma}

\def\be{\begin{equation}}
\def\ee{\end{equation}}

\def\bea{\begin{eqnarray}}
\def\eea{\end{eqnarray}}

\def\ba{\begin{array}}
\def\ea{\end{array}}

\def\bc{\begin{center}}
\def\ec{\end{center}}

\def\bl{\begin{flushleft}}
\def\el{\end{flushleft}}

\def\br{\begin{flushright}}
\def\er{\end{flushright}}

\def\bi{\begin{itemize}}
\def\ei{\end{itemize}}

\def\bt{\begin{tabular}}
\def\et{\end{tabular}}

\newtheorem{question}{Question}
\def\bq{\begin{question}}
\def\eq{\end{question}}

\newtheorem{definition}{Def}
\def\bd{\begin{definition}}
\def\ed{\end{definition}}

\newtheorem{answer}{Answer}
\def\ban{\begin{answer}}
\def\ean{\end{answer}}

\newtheorem{possibleanswer}{Possible answer}
\def\bpa{\begin{possibleanswer}\normalfont}
\def\epa{\end{possibleanswer}}

\newtheorem{theorem}{Theorem}
\def\bth{\begin{theorem}}
\def\eth{\end{theorem}}

\begin{document}

\begin{titlepage}
\vspace{5cm}

\vspace{2cm}

\begin{center}
\bf \Large{Weyl and Ricci gauging from the coset construction}

\end{center}

\begin{center}
{\textsc {Georgios K. Karananas, Alexander Monin}}
\end{center}

\begin{center}
{\it Institut de Th\'eorie des Ph\'enom\`enes Physiques, \\
\'Ecole Polytechnique F\'ed\'erale de Lausanne, \\ 
CH-1015, Lausanne, Switzerland}
\end{center}

\begin{center}
\texttt{\small georgios.karananas@epfl.ch} \\
\texttt{\small alexander.monin@epfl.ch} 
\end{center}

\vspace{2cm}

\begin{abstract}

In this paper we demonstrate how, using the coset construction, a theory can be systematically made Weyl invariant by gauging the scale symmetry. We show that an analog of the inverse Higgs constraint allows the elimination of the Weyl vector (gauge) field in favor of curvatures. We extend the procedure -- previously coined Ricci gauging -- and discuss its subtlety for the case of theories with higher derivatives of conformally variant fields.

\end{abstract}

\end{titlepage}

\section{Introduction}

Theories possessing scale and conformal invariance (see, for example,~\cite{DiFrancesco:1997nk, Rychkov:2016iqz}) constitute a very interesting subject for investigations. They appear ubiquitously for describing physical systems, whenever a separation of scales exists. The presence of these symmetries restricts sufficiently the dynamics, so that more properties of the system can be inferred. In some cases, the theory can even be solved completely. Thus, they give an important handle on quantum field theory in general (for recent progress see~\cite{Komargodski:2011vj,Komargodski:2011xv,Luty:2012ww,Dymarsky:2013pqa, ElShowk:2012ht}).

A very powerful tool for studying these theories is coupling them to a nondynamical metric~\cite{Komargodski:2011vj,Luty:2012ww,Gretsch:2013ooa}. In an even more general setup, all the couplings are considered as background sources~\cite{Baume:2014rla}. It is usually assumed that a conformally invariant theory can be embedded in a curved background in a Weyl invariant manner. It is necessary that a theory be conformal in flat spacetime, in order to couple it to gravity in a Weyl-invariant way. It has been shown that the condition becomes sufficient, only if actions with at most one derivative of conformally variant fields are considered~\cite{Iorio:1996ad}. However, to the best of our knowledge, there is no proof for the condition to be sufficient in general.

The authors of~\cite{Iorio:1996ad} proceed as follows. Given a scale-invariant theory in flat spacetime, it can be made Weyl invariant by gauging dilatations with the help of an additional field $A _ \m$. It so happens that the Weyl variation of a certain combination of the gauge field\footnote{Throughout this paper, greek letters $(\kappa,\lambda,\ldots)$ are reserved for spacetime indices. The metric-compatible covariant derivative $\nabla_\m$, as well as the Christoffel symbols $\G_{\m\n}^\lambda$ are defined in Appendix~\ref{Christoffel}.}
\be
\Theta _ {\m \n} = \nabla _ \m A _ \n - A _ \m A _ \n + \f {1} {2} g _ {\m \n} A ^ \s A _ \s \ ,
\label{tensor_A}
\ee
where $\nabla$ denotes the standard covariant derivative (see Appendix \ref{Christoffel}) and $g_{\m\n}$ the metric,
does not depend on $A _ \m$. It is proportional to the variation of the Schouten tensor
\be
S _ {\m \n} = R _ {\m \n} - \f {R} {2 (n-1)} g _ {\m \n} \ ,
\label{tensor_R}
\ee
with the following convention for the curvatures
\be
\label{ricci-conv}
R = R ^ \m _ \m, ~~~ R _ {\m \n} = \d ^ \lambda _ \s R ^ {~~~\s}_ { \lambda \m ~ \n }  ~~~\text{and}~~~ R ^ {~~~\s}_ { \lambda \m ~ \n } = \p _ \lambda \G _ {\m \n }^ \s - \p _ \m \G _ {\lambda \n }^ \s + 
\G _ {\lambda \rho} ^ \s \G _ {\m \n} ^ \rho - \G _ {\m \rho} ^ \s \G _ {\lambda \n} ^ \rho \ .
\ee
Therefore, if the gauge field enters the Lagrangian only in the combination~\eqref{tensor_A}, it is possible to trade it for the expression in~\eqref{tensor_R}, leaving all the symmetries intact. As a result, the theory becomes Weyl invariant and no additional degrees of freedom are introduced. The authors call this procedure ``Ricci gauging''. Lastly, they prove that for a theory without higher derivatives of conformally variant fields, the described Weyl gauging leads necessarily to the appearance of the tensor~\eqref{tensor_A}, provided the theory is conformal. Consequently, these theories can be made Weyl invariant when coupled to gravity.

The tensor composed of the Weyl gauge field and possessing the transformation properties of~\eqref{tensor_A} can be found by trial and error, but a systematic recipe is missing in the paper. It should also be noted that contrary to the standard gauging of internal symmetries, the Weyl gauge field $A_\m$ appears in the covariant derivative not only with the operator of dilatations, but with the generators of Lorentz transformations as well. This happens because scale invariance is a spacetime symmetry (which does not commute with spacetime translations).

In this paper, we show how Weyl and Ricci gauging can be carried out in a more systematic way. To achieve that, we employ the coset construction, which was first introduced to analyze nonlinearly realized (broken) internal symmetries~\cite{Coleman:1969sm}. Later, it was extended to include spacetime symmetries as well~\cite{Salam:1970qk, Ivanov:1975zq}. The gauging within this framework corresponds to introducing the covariant derivative in the standard way, i.e. with gauge fields corresponding to each generator of the symmetry group.
This method was successfully used in~\cite{Ivanov:1981wn} to formulate gravity as a gauge theory of the Poincar\'e group and in~\cite{Delacretaz:2014oxa} to study the dynamics of relativistic spinning objects.

The main feature of the nonlinear realization of spacetime symmetries (as compared to internal ones) is the counting of degrees of freedom. For the case of internal symmetries, the number of Goldstone modes is always equal to the number of broken generators. For spacetime symmetries, this need not be the case, since it is not rare that a smaller number of Goldstone bosons is enough to realize a symmetry breaking pattern. This happens because the fluctuations produced by the action of all broken generators on the vacuum are not independent. 

From the physical point of view, this phenomenon manifests itself through the equations of motion, when at low energies certain modes may become gapped and, therefore, can be explicitly integrated out. From a more formal perspective, it can be understood with the help of the inverse Higgs mechanism, which consists of imposing covariant (consistent with all symmetries) constraints on the system and solving them algebraically, thus, reducing the number of necessary fields~\cite{Ivanov:1975zq,Low:2001bw,Endlich:2013vfa,Brauner:2014aha}.

In the present paper, we show that both the form of the covariant derivative with the Weyl gauge field and the relation between~\eqref{tensor_A} and~\eqref{tensor_R}, can be obtained by the analog of the inverse Higgs constraints. We use the word analog, because what is usually called inverse Higgs mechanism is a constraint that can be solved algebraically with respect to a certain field (or fields). In our case (for a theory without torsion) we find a constraint that leads to the relation 
\be
(n-2)\Theta_{\m\n} \simeq S_{\m\n} \ .
\label{Ricci_Weyl}
\ee
The reason we use the symbol ``$\simeq$'', is because we want to stress that the above expression is not an equality in the sense that the field $A _ \m$ can be expressed in terms of the metric; it is clear that this equation cannot be solved algebraically. Rather, what we imply is that the combination on the left-hand side of~\eqref{Ricci_Weyl} transforms identically to the one on the right-hand side. Therefore, it can be substituted by the latter in a consistent with all the symmetries way. We also show that once the requirement of having a torsionless theory is relaxed, $A_\m$ is found to be equal to one of the irreducible components of the torsion tensor.

This paper is organized as follows. In Sec.~\ref{sec:coset_constr}, we review the coset construction for space-time symmetries. In Sec.~\ref{sec:local_scale_trans}, we gauge scale transformations and obtain the relation between $\Theta_{\m\n}$ and $S_{\m\n}$ . In Sec.~\ref{sec:examples}, we demonstrate how Ricci gauging works by considering two examples. The first one is the purely gravitational Weyl square theory in four dimensions, and the second one is the $n$-dimensional generalization of the Riegert theory. In Sec.~\ref{sec:altern}, we discuss how Weyl gauging can take place if torsion is present in the theory. Section~\ref{sec:conslusions}, contains the conclusions.  Details on how the Christoffel symbols and the covariant derivative are defined can be found in Appendix~\ref{Christoffel}. The conformal algebra, as well as an outline of the procedure that can be followed for gauging the full conformal group are presented in Appendix~\ref{conf_group}. The irreducible decomposition of the torsion tensor in a $n$ dimensional spacetime is carried out in Appendix~\ref{irred_decomp}.

\section{Coset construction} 
\label{sec:coset_constr}

In this section, we briefly describe the coset construction for spacetime symmetries. Following the standard approach for internal symmetries in flat spacetime~\cite{Coleman:1969sm}, we separate the generators of a symmetry group G (with algebra $\mathfrak{g}$) into two sets. One
contains the symmetry generators that are linearly realized, whereas the other contains the ones which are nonlinearly realized. The broken generators will be denoted by $T$, while the unbroken ones by $t$ and $P$. Notice that we have separated the momenta $P$ from the rest of unbroken generators, for there is a difference between the coset construction for internal and spacetime symmetries already at the stage of choosing the coset representative 
$\Omega$. Namely, in the latter case, one includes the momenta operators in the coset 
\be
\Omega = e ^ {i P x} e ^ {i \pi (x) T} \ .
\label{coset_rep_x}
\ee
Provided that the commutation relations for the generators have the following schematic form
\be
\l [ t, P \r ] = i P \ \ \ \text{and} \ \ \  \l [ t, T \r ] = i T \ ,
\ee
computing the Maurer-Cartan form 
\be
\Omega ^ {-1} \p _ \m \Omega = i e _ \m ^ \n P _ \n + i \nabla _ \m \pi \, T + i \omega _ \m \, t \ ,
\label{MC_flat}
\ee
produces the fields with the following transformation properties (the sum over the indices of the generators is tacitly assumed)
\be
\begin{aligned}
\nabla _ \m \pi ' T & =   \bar h \, \nabla _ \m \pi T \, \bar h ^ {-1} \ ,  \\
\omega _ \m ' T & =   \bar h \, \omega _ \m t \, \bar h ^ {-1} + \bar h \p _ \m \bar h ^ {-1}\ , \\
e _ \m ^ {' \n} P _ \n & =  \bar h \, e _ \m ^ \n P _ \n \, \bar h ^ {-1}\  .
\label{coset_trans_h}
\end{aligned}
\ee
Here $\bar h (x, g)$ is a certain function that can be found from
\be
\Omega '  = g \, \Omega \, \bar h ^ {-1} (x, g) \ .
\ee
As a result, we have the necessary building blocks to analyze a system with spontaneously broken symmetries. For example, any $H$-invariant function of $\nabla _ \m \pi$ would produce a Lagrangian which is ``secretly'' $G$ invariant, if one also uses $e_ \m ^ \n$ to build an invariant measure. Similarly, the connection $\omega _ \m t$ can be used to construct higher derivative terms and/or coupling to matter fields.

It should be noted that since the quantities $\nabla _ \m \pi$ transform in a covariant way, it is consistent to set them to zero. In certain cases~\cite{Ivanov:1975zq,Low:2001bw}, the explicit form of some of the $\nabla _ \m \pi$ allows for an algebraic solution. In this case, a subset of Goldstone modes gets expressed in terms of derivatives of other fields. This is called the inverse Higgs mechanism.

The reason that the momentum operators are included in the coset, and moreover they appear as a separate exponent, is that they belong to the unbroken subgroup $H$ (with algebra~$\mathfrak{h}$). Therefore, they should be realized linearly on the fields. It proves useful, especially having in mind further gauging of the Poincar\'e group, to introduce the auxiliary fields $y$ that multiply the momenta in the exponent~\eqref{coset_rep_x}. In doing so, one arrives to a situation similar to the one corresponding to the breaking of internal symmetries. Namely, in this case the expression
\be
\Omega = e ^ {i P y(x)} e ^ {i \pi (x) T} \ ,
\label{coset_rep_y}
\ee
looks exactly like the would-be coset representative corresponding to the internal symmetries, with $P$ and $T$ broken spontaneously. We can now perform the gauging in the usual way, simply by promoting the partial derivative in~\eqref{MC_flat} to a covariant one.

\section{Local scale transformations}
\label{sec:local_scale_trans}

The way to introduce gravity within this framework is to promote Poincar\'e transformations to be local ones, and at the same time to demand that the theory be invariant under general coordinate transformations (diffeomorphisms)~\cite{Ivanov:1981wn}. Our goal is to obtain a Weyl-invariant theory, consequently, we will gauge scale transformations as well. In this case, the coset representative does not contain generators other than the momenta\footnote{We use capital latin letters $(A,B,\dots)$ for internal (Lorentz) indices. The Minkowski metric is mostly minus, i.e. $\eta _ {AB} = \mathrm{diag} \l ( 1,-1,-1,\ldots \r )$.}
\be
\Omega = e ^ {i P _ A y ^ A} \ .
\ee
Therefore, the Maurer-Cartan form becomes
\be
\Omega ^ {-1} \l ( \p _ \m + i \tilde e ^ A _ \m P _ A + \f {i} {2} \tilde \omega ^ {AB} _ \m J _ {AB} + i \tilde A _ \m D \r ) \Omega = 
i e _ \m ^ A P _ A + \f {i} {2} \omega ^ {AB} _ \m J _ {AB} + i A _ \m D \ ,
\ee
where $\tilde e_\m^A$ , $\tilde \omega_\m^{AB}$ and $\tilde A_\m$ are gauge fields corresponding to translations, Lorentz rotations and dilatations respectively, while their counterparts without the tilde can be viewed as the fields in the unitary gauge. It is straightforward to check that the transformation properties of $e _ \m ^ A$ allow for it to be interpreted as a vielbein that is used to mix spacetime and Lorentz indices, to define the metric $g_{\m\n}=e_\m^A e_\n^B \eta_{AB}$, and to construct the diffeomorphism-invariant measure\footnote{If we do not require that the theory be invariant under the full group of diffeomorphisms, then the construction of the invariant measure is not necessary. For example one may be interested in theories invariant only with respect to volume preserving diffeomorphisms (or transverse diffeomorphisms, see for example~\cite{Buchmuller:1988wx, *Alvarez:2006uu, *Blas:2011ac} and references therein). In this case, the theory is invariant only under the subgroup of coordinate transformations with Jacobian equal to unity, thus we can allow for the presence of arbitrary powers of the vielbein determinant.}
\be
d ^ n x \, \det e_{\m}^A\equiv d ^ n x \, \det e \ .
\label{detemeasure}
\ee
$\omega ^ {A B} _ \m$ in turn is interpreted as the spin connection. Indeed, using the analog of~\eqref{coset_trans_h}
\be
\Omega '  =  g \Omega h ^ {-1} (y,g), ~~ \text{with} ~~ h = e ^ {- i t \a (y,g)} \in H = SO (n-1,1) \times \mathbb R \ ,
\ee
and the commutation relations presented in Appendix~\ref{conf_group}, one finds the transformation properties of the gauge fields

\begin{table}[H]
\centering
\bt{c | ccc}
$ $ &$e ^ {'A} _ {\m}$ & $\omega ^ {' AB} _ {\m} $ & $A' _ {\m}$ \\
\hline
$J$ & $ e _ {\m} ^ B \Lambda _ B ^ {~A}$ & $\omega ^ {CD} _ {\m} \Lambda _ {C} ^ {~A} \Lambda _ {D} ^ {~B}+\l(\Lambda\p_\m \Lambda^{-1}\r)^{AB}$ & $A _ {\m}$ \\
$D$ & $ e ^ {-\a} e _ {\m} ^ A $ & $\omega ^ {AB} _ {\m}$ & $A _ {\m} + \p _ \m \a $
\et
\end{table}

The transformations of $e _ \m ^ A$, $\omega _ \m ^ {AB}$ and $A_ \m$ are precisely the ones for the vielbein, spin connection and the Weyl gauge field. According to the rules of the coset construction, the covariant derivative of a matter field $\psi$ is given by
\be
D _ A \psi  = e ^ \m _ A \l ( \p _ \m + \f {i} {2} \omega ^ {A B} _ \m J _ {AB} + i A _ \m D \r ) \psi \ ,
\label{coset_covd}
\ee
where at this stage, $\omega _ \m ^ {AB}$ are considered as independent degrees of freedom. To express them in terms of the vielbein, as it is usually done for torsionless gravity, we should impose some constraints. We will be back to this point shortly.

It is clear that by analogy with gauge field theories, one can construct field strength tensors corresponding to shifts, Lorentz and scale transformations
\begin{align}
\centering
\label{tensor-dil-1}
e _{ \m \n } ^ A & =  \p _ \m e _ \n ^ A - \p _ \n e _ \m ^ A - \omega _ {\m B} ^ {A}  e _ \n ^B + \omega _ {\n B} ^ {A}  e _ \m ^B +
A _ \m E _ \n ^ A - A _ \n E _ \m ^ A\ ,  \\
\label{tensor-dil-2}
\omega ^ {AB} _ {\m \n} & =  \p _ \m \omega _ \n ^ {AB} - \p _ \n \omega _ \m ^ {AB} 
- \omega _ {\m C} ^ {A} \omega _ \n ^{CB} + \omega _ {\n C} ^ {A} \omega _ \m ^{CB}\ ,  \\
\label{tensor-dil-3}
A _ {\m \n} & =  \p _ \m A_ \n - \p _ \n A _ \m \ , 
\end{align}
that transform covariantly

\begin{table}[H]
\centering
\bt{c | cccc}
$ $ &$e ^ {'A} _ {\m \n}$ & $\omega ^ {' AB} _ {\m \n} $ & $A' _ {\m \n}$ \\
\hline
$J$ & $ e _ {\m \n} ^ B \Lambda _ B ^ {~A}$ & $\omega ^ {CD} _ {\m \n} \Lambda _ {C} ^ {~A} \Lambda _ {D} ^ {~B}$ & $A _ {\m \n}$ \\
$D$ & $ e ^ {-\a} e _ {\m \n} ^ A $ & $\omega ^ {AB} _ {\m \n}$ & $A _ {\m \n}$
\et
\end{table}

As we discussed in the Introduction, it is consistent to set to zero any covariant quantity. However, not all of them can be solved algebraically. The ones that can be solved are those that get a contribution from the commutator of the momentum and another generator.\footnote{Compare this with the standard inverse Higgs constraint condition: if  $\l [ P _ \m, X \r ] \supset X '$, then the Goldstone corresponding to $X$ can be expressed as a derivative of that of $X '$, by solving $\nabla _ \m \pi _ {X'} = 0$.} For the case at hand, the commutators of $P _ A$ and $J _ {AB}$ (see Appendix~\ref{conf_group}) suggest that the constraints 
\be
e _ {\m \n} ^ A = 0 \ ,
\label{shift_IH}
\ee
could be solved. Indeed, the solution takes the form
\be
\label{spin-con3}
\omega ^ {AB} _ \m = \bar \omega^{AB}_\m + \d\omega^{AB}_\m \ ,
\ee
where we defined
\be
\label{spin-con4}
\bar \omega^{AB}_\m=-\frac{1}{2}\l[e ^ {\n A} \l ( \p _ \m e _ \n ^ B - \p _ \n e _ \m ^ B \r )  - e ^ {\n B} \l ( \p _ \m e _ \n ^ A - \p _ \n e _ \m ^ A \r ) 
- e _ {C \m} e ^ {\n A} e ^ {\lambda B} \l ( \p _ \n e _ \lambda ^ C - \p _ \lambda e _ \n ^ C \r ) \r] \ ,
\ee
which is the standard spin connection for a torsionless theory and
\be
\label{spin-con5}
\d\omega_\m^{AB}= I_{\m \n}^{A B} A^ \n \ , \ \ \  I_{\m \n}^{A B}= e_\n^A e_\m^B-e_\n^B e_\m^A \ .
\ee
Plugging the expression for $\omega$ to the definition of the covariant derivative~\eqref{coset_covd}, we find that it can rewritten as follows
\be
D _ A \psi  = e ^ \m _ A \l ( \p _ \m + \f {i} {2} \bar \omega ^ {A B} _ \m J _ {AB} - i e ^ A _ \m e ^ B _ \n A ^ \n J _ {A B} + i A _ \m D \r ) \psi \ .
\label{coset_covd_e}
\ee
In particular, for a vector field $V ^ A$ with scaling dimension $\D _ V$, we get
\be
e ^ B _ \m D _ B V ^ A = \p _ \m V ^ A - \bar \omega ^ A _ {\m B} V ^ B + ( e _ \m ^ A e _ B ^ {\n} - e ^ {\n A} e _ {\m B} )V ^ B A _ \n - \D _ V A _ \m V ^ A \ .
\label{covd_vector}
\ee
We can clearly see now the reason why the Weyl gauge field ``couples'' to  spin as well. Using the Christoffel symbols defined in Appendix~\ref{Christoffel}, one can show that the expression for the covariant derivative~\eqref{coset_covd_e} coincides with the one used in~\cite{Iorio:1996ad}.

Notice that the field strength tensor corresponding to shifts is not the only covariant structure. Even though imposing another constraint is not in the spirit of the standard inverse Higgs mechanism, it can be done consistently.\footnote{For pure Poincar\'e invariance, no additional constraint is usually imposed, since there are no candidates for elimination, provided one wants to obtain dynamical gravity.} The gauge field $\omega _ \m ^ {AB}$ depends on $A_ \m$; therefore, we may hope to relate certain structure depending on this vector to a tensor that depends only on the vielbein.

Plugging the expression (\ref{spin-con3}) to the formula (\ref{tensor-dil-2}), we get
\be
\omega ^ {AB} _ {\m \n} = \bar \omega ^ {AB} _ {\m \n} + \d \omega ^ {AB} _ {\m \n},
\label{omega_full}
\ee
with
\be
\label{barR}
\bar \omega _{\m\n}^{AB} = \p _ \m \bar\omega _ \n ^ {A B} - \p _ \n \bar\omega _ \m ^ {A B} 
-\bar \omega _ {\m C} ^ {A} \bar\omega _ \n ^{CB} +\bar \omega _ {\n C} ^ {A} \bar\omega _ \m ^{CB} \ ,
\ee
and
\be
\begin{aligned}
\label{del_omega}
\d \omega ^ {AB} _ {\m \n} & = I^{AB}_{\n\lambda}\nabla_\m A^\lambda -I^{AB}_{\m\lambda}\nabla_\n A^\lambda 
+\l(e^A_\m e^B_\n-e^B_\m e^A_\n\r)A^2 \\
&+
\l(e^A_\n A^B-e^B_\n A^A\r)A_\m - \l(e^A_\m A^B-e^B_\m A^A\r)A_\n \ , 
\end{aligned}
\ee
where we used the vielbein to manipulate the indices of $A _ \m$, so that $A^2=A_B A^B=A_\m A^\m$.

None of the constraints imposed on $\omega ^ {A B} _ {\m \n}$, although consistent with its transformation properties, can be solved algebraically with respect to $A _ \m$. Nevertheless, imposing
\be
\omega _ {\m \n} + \omega _ {\n \m} \simeq 0 \ , \ \ \  \text{with} \ \ \  \omega _ {\m \n} \equiv \omega ^ {A B} _ {\m \s} e ^ {\s} _ B e _ {\n A} \ ,
\label{omega_constr}
\ee
and using~\eqref{omega_full} leads to~\eqref{Ricci_Weyl}, which coincides with the expression obtained in~\cite{Iorio:1996ad}, except that we use a different convention for the Riemann curvature tensor, see~\eqref{ricci-conv}.

The substitution $S _ {\m \n}$ for $\Theta _ {\m \n}$ is similar in the spirit to the standard inverse Higgs mechanism, according to which, certain degrees of freedom are not needed to realize a symmetry breaking pattern and as a result, they can be eliminated. Note, however, that the opposite substitution is not legitimate (at least not for arbitrary field configurations), since the Schouten tensor is subject to the Bianchi identity
\be
\nabla ^ \m S _ {\m \n} - \nabla _ \n S = 0 \ ,
\label{Bianchi_Schouten}
\ee
which is not satisfied by $\Theta _ {\m \n}$.

In~\cite{Iorio:1996ad}, it was shown that the substitution~\eqref{Ricci_Weyl} can always be made for conformal (in flat spacetime) theories with at most one derivative of conformally variant fields. In this case, the invariance under Weyl rescalings does not require the introduction of extra degrees of freedom, since the inhomogeneous pieces of the transformation that appear in the derivatives can be compensated for by curvature terms. 

It should also be noted that the constraint~\eqref{omega_constr} taken as an equality, only implies the equivalence between the Schouten tensor $S _ {\m \n}$ and the symmetric part of $\Theta _ {\m \n}$. However, in a weaker sense (that is, equivalence of the transformation properties), it is possible to relate 
$S _ {\m \n}$  to the full $\Theta_{\m\n}$. In fact, the antisymmetric part is given by
\be
2 \Theta _ {\m \n} ^ {\text{anti}} = \nabla _ \m A _ \n - \nabla _ \n A _ \m  = A _  {\m \n} \ ,
\ee
which is invariant under Weyl transformations and can be safely added to $\Theta _ {\m \n} ^ {\text{sym}}$, resulting in~\eqref{Ricci_Weyl}.

\section{Examples}
\label{sec:examples}

\subsection{Weyl tensor}

As a first example, we build the Weyl-invariant action for pure gravity in a four dimensional spacetime, without coupling to matter.  After imposing the constraint $e ^ A _ {\m \n} = 0$, we are left with three objects: two Weyl-invariant curvatures $\omega ^ {A B} _ {\m \n}$ and $A _ {\m \n}$, and the Weyl covariant vielbein $e ^ A _ \m$. In order to account for the noninvariance of the measure $d ^ 4 x \, \det e$,  it should be multiplied four times by the inverted vielbein
\be
\int d ^ 4 x \, \det e \, e ^ \m _ A \, e ^ \n _ B \, e ^ \lambda _ C \, e ^ \s _ D \ .
\label{measure_W}
\ee
The lowest-order (in derivatives) diffeomorphism-invariant action, which also respects the gauged scale and Poincar\'e symmetries 
can be obtained by all possible contractions of (\ref{measure_W}) with
\be
A _ {\m \n} A _ {\lambda \s} ~~~\text{and} ~~~\omega ^ {AB} _ {\m \n} \omega ^ {AB} _ {\lambda \s} \ .
\ee
The first term leads to the following obviously Weyl-invariant action (we do not assume parity invariance)
\be
S _ 1 = \int d ^ 4 x \, \det e \, \l ( c _ 1 A _ {\m \n} A ^ {\m \n} + c _ 2 \ep ^ {\m \n \lambda \s} A _ {\m \n} A_ {\lambda\s} \r ),
\label{A_W_inv}
\ee
 with $c_1$ and $c_2$ being constants and $\ep ^ {\m \n \lambda \s} = e ^ \m _ A \, e ^ \n _ B \, e ^ \lambda _ C \, e ^ \s \ep ^ {ABCD}$. The contractions with $\omega ^ {AB} _ {\m \n} \omega ^ {AB} _ {\lambda \s}$ can be simplified once the constraint (\ref{omega_constr}) is imposed. The antisymmetric part of $\omega _ {\m \n}$ from (\ref{omega_constr}) is proportional to $\p _ \m A _ \m - \p _ \m A _ \m$, which already has been taken into account in~\eqref{A_W_inv}. We may thus consider only configurations with $\omega ^ {A B} _ {\m \n} e ^ \n _ B = 0$. As a result, the only possible contractions are the following
\be
\label{conf-inv-act2}
\ep^{IJKL}e^\m_I e^\n_J e^\rho_K e^\s_L\ep_{ABCD}~\omega_{\m\n}^{AB}\omega_{\rho\s}^{CD} \ , 
\ee
and
\be
\label{conf-inv-act3}
\ep^{IJKL}e^\m_I e^\n_J  e^\rho_K e^\s_L~\eta_{AC}\eta_{BD}\ , \ \ \ \ep^{IJKL}e^\m_I e^\n_J   E^\rho_A E^\s_B~\eta_{KC}\eta_{LD} \omega_{\m\n}^{AB}\omega_{\rho\s}^{CD} \ .
\ee 
Simplifying these expressions, leads to
\be
S _ 2 = \int d ^ 4 x \, \det e \, \l ( c _ 3 W _ {\m \n \lambda \s} W ^ {\m \n \lambda \s} + 
c _ 4 \ep ^ {\kappa \rho \lambda \s} W _ {\m \n \kappa \rho}  W ^ {\m \n} _ {~~~\lambda \s} \r ),
\ee
where $W _ {\m \n \lambda \s}$ is the Weyl tensor and $c_3$, $c_4$ are constants.

\subsection{Higher derivative action}

In this section, we wish to get a better grasp on the range of applicability of Ricci gauging. To be more precise, we want to understand whether or not the presence of more than one derivative of a conformally variant field constitutes an obstruction in the Ricci gauging, as claimed in~\cite{O'Raifeartaigh:1996hf}. To achieve that, we consider a theory with a higher number of derivatives of a scalar field, namely, a conformally invariant theory in an $n$-dimensional flat spacetime given by the following action 
\be
S_{\Box^2}=\int d^n x ( \Box \phi ) ^ 2 \ .
\ee
According to the coset construction described previously, we introduce the covariant derivative~\eqref{coset_covd_e} for the field $\phi$ in the following way
\be
D _ A \phi = e ^ \m _ A \l ( \nabla _ \m \phi - \D A _ \m \phi \r ) \ ,
\ee
where $\D = \f {n} {2} - 2$ is the scaling dimension of $\phi$. Therefore,
\be
e ^ B _ \m D _ B D _ A \phi = \p _ \m D _ A \phi - \bar \omega _ {\m A } ^ {~~B} D _ B \phi + ( e _ {\m  A} e ^ {\n B} - e ^ {\n} _ A e _ {\m} ^ B ) D _ B \phi A _ \n - ( \D + 1 ) A _ \m D _ A \phi \ ,
\ee
where we used the fact that the scaling dimension of $D _ A \phi$ is equal to $\D+1$. As a result, the following substitution
\be
\Box \phi \to D _ A D ^ A \phi = \nabla ^ 2 \phi + 2 A _ \m \nabla ^ \m \phi - \l ( \f {n} {2} - 2 \r ) \l ( \nabla _ \m A ^ \m + \f {n} {2} A ^ \m A _ \m \r ) \ ,
\ee
where $\nabla ^ 2 = g ^ {\m \n} \nabla _ \m \nabla _ \n$, leads to the Weyl-invariant action
\be
S _ {\Box ^ 2} = \int d ^ n x \, \det e ( D _ A D ^ A \phi ) ^ 2 \ .
\label{Lagr_box_2}
\ee
The question we would like to address now is whether it is possible to use Ricci gauging [or, equivalently, the weak form of the constraint~\eqref{Ricci_Weyl}] to completely get rid of the field $A _ \m$. Lengthy but straightforward calculations lead to 
\be
\begin{aligned}
S_ {\Box ^ 2} & = \int d^4x \det e~\left\{\vphantom{\frac{A}{B}}( \nabla ^ 2 \phi ) ^ 2 - \l [ 4 \Theta _ {\m \n} - (n-2) \Theta g _ {\m \n} \r ] \nabla ^ \m \phi \nabla ^ \n \phi\right. \\
&\left. -  \phi ^ 2 \l [ \f {n-4} {2} \nabla ^ 2 \Theta + (n-4) \Theta _ {\m \n} \Theta ^ {\m \n} - \f {n (n-4)} {4} \Theta ^ 2 \r ]\right.  \\
&\left.- \phi ^ 2 (n - 4) A ^ \n \nabla ^ \m \l (  \Theta _ {\m \n} - g _{\m\n} \Theta \r ) \vphantom{\frac{A}{B}}\right\}\ ,
\end{aligned}
\ee
where $\Theta = \Theta ^ \m _ \m$. Notice that the dependence of the action on $A _ \m$ for $n= 4$ is only through the tensor $\Theta _ {\m \n}$ and Ricci gauging can be used without any trouble. Although, for general $n$, there is an explicit $A _ \m$ dependence in the last term, it is clear that after the substitution (we assume $n \neq 2$)
\be
\Theta _ {\m \n} \to \f {1} {n-2} S _ {\m \n} \ ,
\ee
this term drops out by virtue of the Bianchi identity~\eqref{Bianchi_Schouten}. Therefore, it is shown that the theory given by the Lagrangian (\ref{Lagr_box_2}) can be Ricci gauged in an arbitrary (not equal to two) number of dimensions. The resulting action can be written in the following form 
\be
S _ {\Box ^ 2} = \int d ^ n x \sqrt {g} \phi \mc Q (g)  \phi,
\ee
with
\be
\begin{aligned}
\mc Q (g) = \nabla ^ 2 + \nabla ^ \m  \l[\l(\frac{4} {n-2} S _ {\m \n} -  g _ {\m \n}S \r ) \nabla ^ \n\r] - \frac{n-4} {2 (n-2)} \nabla ^ 2 S - \frac{n-4}{(n-2)^2} S _ {\m \n} S ^ {\m \n} + \frac{n (n-4)}{4 (n-2)^2} S ^ 2 \ , 
\label{pan-rieg}
\end{aligned}
\ee
being the Paneitz operator~\cite{Paneitz:1983_2008}, which is the Weyl covariant generalization of $\Box ^ 2$, see also Appendix~\ref{riegert}.

\section{Torsionful theory}
\label{sec:altern}

The field strength corresponding to shifts $e^ A _ {\m \n}$ and the generalized spin connection $\omega ^ {A B} _ \m$ have the same symmetry properties; therefore, they have equal number of independent components. This is the reason why we were able to solve the inverse Higgs constraint~\eqref{shift_IH} with respect to the  $\omega ^ {A B} _ \m$ and express it in terms of the vielbein and the Weyl vector field $A_\m$. This way, we built a Weyl-invariant torsionless theory. Here we look for an alternative solution to this constraint. 

In order to understand what the possible solutions might be, we should analyze the structure of irreducible representations of $e ^ A _ {\m \n}$, since they can be set to zero independently. Any tensor that possesses the symmetries of the quantity $e ^ A _ {\m \n}$, admits the following decomposition in an $n$-dimensional spacetime (see also Appendix~\ref{irred_decomp}). A vector,
\be
\varepsilon _ \m =e ^ \n _ A e ^ A _ {\m \n} \ ,
\label{vect}
\ee
a completely antisymmetric tensor
\be
\mathscr A ^ { \s _ 1 \dots \s _ {n-3}} = \frac{1}{n} \,\ep^{\s_1\s_2\cdots \s_{n-3}\m\n\lambda}e_{\lambda A}\, e_{\m\n}^A \ ,
\label{antisym}
\ee
and a traceless tensor with mixed symmetries
\be
\mathscr E_{\m\n}^A=e_{\m\n}^A-\frac{3}{2(n-1)}\l(\varepsilon _\m e_\n^A- \varepsilon_\n e^A_\m \r)-\frac{1}{2}e^{\lambda A}\l(e^B_{\lambda\m} e_{\n B}-e^B_{\lambda\n} e_{\m B}\r) \ .
\label{traceless}
\ee
Written in this form, the constraints~\eqref{shift_IH}, make it clear that~\eqref{antisym} and~\eqref{traceless} can only be solved with respect to their counterparts contained in $\omega ^ {A B} _ \m$. However, for the vector part~\eqref{vect} there are two options. The first one, which has been chosen in the previous section, is to eliminate the vectorial part of the spin connection. The second one is to solve the constraint with respect to $A _ \m$, keeping $\omega ^ {AB} _ \m e ^ \m _ B$ undetermined, which yields a torsionful theory.

We see from~\eqref{tensor-dil-1} that 
\be
\label{tors-A-1}
A_\m e_\n^A-A_\n e_\m^A= -T_{\m\n}^A \ ,
\ee
where the torsion tensor $T_{\m\n}^A$ is defined as
\be
\label{tors-tens-1}
T_{\m\n}^A\equiv \p _ \m e _ \n ^ A - \p _ \n e _ \m ^ A - \omega _ {\m B} ^ {A}  e _ \n ^B + \omega _ {\n B} ^ {A}  e _ \m ^B \ , 
\ee
Tracing~\eqref{tors-A-1}, we obtain
\be
\label{tors-A-2}
A_\m=-\frac{1}{n-1} \upsilon_\m \ ,
\ee
where we denoted with  $\upsilon_\m$ the torsion vector
\be
\label{tors-A-4}
\upsilon_\m= e^\n_AT_{\m\n}^A= e^\n_A\l (\p _ \m e _ \n ^ A - \p _ \n e _ \m ^ A + \omega _ {\n B} ^ {A}  e _ \m ^B \r) \ .
\ee

It is straightforward to check that under Weyl rescalings the vector $\upsilon _ \m$ transforms exactly as the Weyl field, i.e. 
\be
\upsilon'_\m=\upsilon_\m-(n-1)\p_\m \alpha \ .
\ee
As a result, once we consider nonvanishing torsion, the degrees of freedom carried by $A_\m$ can be traded for the vector $\upsilon_\m$.

We should mention that certain torsionful theories~\cite{Ho:2011qn,*Ho:2011xf,*Baekler:2010fr} have attracted considerable attention, since they are free from pathologies and have very interesting cosmological phenomenology. In general though, theories in which torsion is propagating are not necessarily ghost and tachyon free; see for example~\cite{Neville:1978bk,*Sezgin:1979zf,*Hayashi:1979wj,*Hayashi:1980ir,*Hayashi:1980qp,*Sezgin:1981xs,*Kuhfuss:1986rb,*Karananas:2014pxa} and references therein.

\section{Conclusions}
\label{sec:conslusions}

In this paper we touched upon the question of whether the conformal invariance of a system in flat spacetime implies that the system can be coupled to gravity in a Weyl-invariant way. We used the prescription of the standard coset construction in order to gauge scale transformations (along with the Poincar\'e group), leading to a Weyl-invariant (in curved spacetime) theory. It was demonstrated that the main ingredient needed for Ricci gauging, namely the relation between the additional gauge field corresponding to the local scale transformations and the Ricci curvature -- first obtained in~\cite{Iorio:1996ad} -- can be extracted from the analog of the inverse Higgs constraint. It is a standard tool in the coset construction applied to the spontaneous breaking of the spacetime symmetries, needed to eliminate unnecessary degrees of freedom. 

In our case, it is used to show that the two structures~\eqref{Ricci_Weyl} transform in the same way, and therefore, whenever the tensor $\Theta _ {\m \n}$ appears in the action, it can be substituted by its counterpart without any contradiction with the underlying symmetries. The answer to the question of whether such a prescription for conformally invariant theories leads to a complete elimination of the gauge field $A _ \m$ does not have a definite answer at the moment and can only be divined. 

We presented a couple of examples of how Ricci gauging works. First, we obtained the Weyl-invariant action for pure gravity in four spacetime dimensions, which is given, as is well known, by the square of the Weyl tensor. Next, we considered a theory with more than one derivative of a scalar field~\eqref{Lagr_box_2}. In a four dimensional spacetime, the Ricci gauging can be straightforwardly employed. However, it so happens that the scaling dimension of the field is zero in this case; thus, the field is actually conformally invariant. Notice that there is no contradiction with~\cite{Iorio:1996ad}, since the condition of having at most one derivative was only imposed on conformally variant fields.

Considering the system in $n \neq 4$, we showed that contrary to what was expected in~\cite{O'Raifeartaigh:1996hf}, Ricci gauging can be applied even for theories with more than one derivative of conformally variant fields. In the example we considered, the procedure turned out to be a little bit subtle. Namely, the Weyl gauged Lagrangian cannot be written as a function depending only on $\Theta _ {\m \n}$, but rather, it also depends explicitly on $A _ \m$. However, this dependence drops out, once Ricci gauging is performed.

Finally, we also presented an alternative way of introducing the Weyl symmetry. We showed, by solving the inverse Higgs constraint, that the role of the gauge field associated with local scale transformations can be played by the vector part of the torsion tensor.

\section*{Acknowledgements}
The work of G.K.K. and A.M. is supported by the Swiss National Science Foundation.

\appendices

\section{Christoffel symbols and covariant derivatives \label{Christoffel}}

The coset construction allows us to write a covariant derivatives for internal symmetries (having introduced the fields $y ^ A$, we have made spacetime translations effectively internal), meaning that it acts only on Lorentz indices  $A,B,\ldots$. However, the procedure does not produce the covariant derivative for fields with spacetime indices or, in particular, for the vielbein. Nevertheless, one can introduce the analog of Christoffel symbols,\footnote{However, one should be careful, since the new symbols depend explicitly on the scaling dimension of fields they act on.} so that the covariant derivative is consistent with interchanging the Lorentz and spacetime indices. Namely, using the vector with scaling dimension $\D _ V$,
\be
V ^ A = e ^ A _ \m V ^ \m \ ,
\ee
one defines
\be
D _ \m V ^ \n = \p _ \m V ^ \n + G ^ \n _ {\m \lambda} V ^ \lambda = e _ A^\n e ^ B _ \m D _ B V ^ A \ .
\ee
Using the expression for $\omega$ from~\eqref{spin-con3}, it is not difficult to show that in this case
\be
G ^ \s _ {\m \n} = \G ^ \s _ {\m \n} + \d G ^ \s _ {\m \n} \ ,
\ee
with $\G$ being the standard Christoffel symbols
\be
\G ^ \s _ {\m \n} = - \l ( e ^ \s _ A \bar \omega ^ {A} _ {\m B} e ^ B _ \n + e ^ \s _ A \p _ {\m} e ^ A _ \n \r ) \ ,
\ee
which are compatible with the metric and thus satisfy
\be
\nabla _ \m V ^ \n \equiv \p _ \m V ^ \n + \G ^ \n _ {\m \lambda} V ^ \lambda = e _ A ^ \n  (\p _ \m V ^ A - \bar \omega ^ A _ {\m B} V ^ B ) \ ,
\ee
 with $\nabla$ the standard covariant derivative. Meanwhile
\be
\d G ^ \s _ {\m \n} = - \D _ V A _ \m \d ^ \s _ \n + A _ \n \d ^ \s _ \m - A ^ \s g _ {\m \n} \ .
\ee
Using the fact that the covariant derivative for a field $V ^ \m$ with scaling dimension $\D _ V + 1$ can be written as
\be
D _ \m V ^ \s = \nabla _ \m V ^ \s +  \underbrace{( A _ \m \d ^ \s _ \n + A _ \n \d ^ \s _ \m - A ^ \s g _ {\m \n})} _ {\d \G ^ \s _ {\m \n}}  V ^ \n - ( \D _ V + 1 ) A _ \m V ^ \s \ ,
\ee 
it is straightforward to show that the covariant derivative $\nabla _ \m$ can be made Weyl covariant, provided all partial derivatives 
are substituted by
\be
\p _ \m \to \p _ \m - \D A _ \m \ ,
\ee
where $\D$ is the scaling dimension of the field the partial derivative $\p _ \m$ acts on. For instance
\be
\p _ \m g _ {\lambda \s} \to \p _ \m g _ {\lambda \s} + 2 A _ \m g _ {\lambda \s}  \ .
\ee

\section{Conformal algebra \label{conf_group}}

The conformal group in $n \neq 2$ dimensions is an extension of the Poincar\'e group. On top of the momenta $P _ A$ (translations) and the Lorentz generators $J_{AB}$, it contains dilatations $D$ and special conformal transformations (SCT) $K_A$, also called conformal boosts. Overall, there are $n(n+1)/2$ generators with the following nonzero commutation relations~\cite{Rychkov:2016iqz,DiFrancesco:1997nk}
\be
\begin{aligned}
\label{conf_cr}
\l [ D, P _ A \r ] & =  - i P _ A \ ,  \\
\l [ J _ {AB}, P _ C \r ] & =  i \l ( \eta _ {BC} P _ A - \eta _ {AC} P _ B \r ) \ ,  \\
\l [ K _ A, P _ B \r ] & =  - 2 i \l ( \eta _ {AB} D + J _ {AB} \r ) \ ,  \\
\l [ D, K _ A \r ] & =  i K _ A\ , \\
\l [ J _ {AB}, J _ {CD} \r ] & =  i \l ( J _ {AD} \eta _ {BC} + J _ {BC} \eta _ {AD} -
J _ {BD} \eta _ {AC} - J _ {AC} \eta _ {BD} \r ) \ ,  \\
\l [ J _ {AB}, K _ C \r ] & =  i \l ( \eta _ {BC} K _ A - \eta _ {AC} K _ B \r ) \ . 
\end{aligned}
\ee

For completeness, let us briefly describe what would happen if the full conformal group was gauged instead of just Poincar\'e and dilatations. It is straightforward to repeat the steps of the coset construction using the commutation relations for the conformal group. This leads to the following transformation rules for the gauge fields

\begin{table}[H]
\centering
\bt{c | cccc}
$ $ &$e ^ {'A} _ {\m}$ & $\omega ^ {' AB} _ {\m} $ & $A' _ {\m}$ & $B ^ {'A} _ {\m}$ \\
\hline
$J$ & $ e _ {\m} ^ B \Lambda _ B ^ {~A}$ & $\omega ^ {CD} _ {\m} \Lambda _ {C} ^ {~A} \Lambda _ {D} ^ {~B}+\l(\Lambda\p_\m \Lambda^{-1}\r)^{AB}$ & $A _ {\m}$ & $ B _ {\m} ^ B \Lambda _ B ^ {~A}$ \\
$D$ & $ e ^ {-\a} e _ {\m} ^ A $ & $\omega ^ {AB} _ {\m}$ & $A _ {\m} + \p _ \m \a $ & $ e ^ {\a} B _ {\m} ^ A $ \\
$K$ & $ e _ {\m} ^ A $ & $\omega _ {\m} ^ {AB} + 2 e _ {\m} ^ {\l [ A \r.} \a ^ {\l. B \r]} $ & $A _ {\m} - 2 e _ {\m} ^ C \a _ C$ & $ B _ {\m} ^ A 
+ \a ^ {A} _ B e _ {\m} ^ B - \omega _ {\m} ^ {AB} \a _ B - \a ^ A A _ {\m} + \p _ \m \a ^ A$
\et
\end{table}

\noindent
Notice that we introduced the new gauge fields $B ^ A_ \m$, associated with  SCT. The corresponding field strengths are found to be
\begin{align}
\centering
\label{tensor-defs1}
E _{ \m \n } ^ A & =  \p _ \m E _ \n ^ A - \p _ \n E _ \m ^ A - \omega _ {\m B} ^ {A}  E _ \n ^B + \omega _ {\n B} ^ {A}  E _ \m ^B +
A _ \m E _ \n ^ A - A _ \n E _ \m ^ A\ ,  \\
\label{tensor-defs2}
\omega ^ {AB} _ {\m \n} & =  \p _ \m \omega _ \n ^ {AB} - \p _ \n \omega _ \m ^ {AB} 
- \omega _ {\m C} ^ {A} \omega _ \n ^{CB} + \omega _ {\n C} ^ {A} \omega _ \m ^{CB}\nn \\
&~~~~~~~~~~~~~+ 
2 \l ( B ^ A _ \m E ^ B _ \n - B ^ A _ \n E ^ B _ \m - B ^ B _ \m E ^ A _ \n + B ^ B _ \n E ^ A _ \m \r )\ ,  \\
\label{tensor-defs3}
A _ {\m \n} & =  \p _ \m A_ \n - \p _ \n A _ \m + 2 \l ( B ^ A _ \m E _ {\n A} - B ^ A _ \n E _ {\m A} \r )\ ,  \\
\label{tensor-defs4}
B _{ \m \n } ^ A & =  \p _ \m B _ \n ^ A - \p _ \n B _ \m ^ A - \omega _ {\m B} ^ {A}  B _ \n ^B + \omega _ {\n B} ^ {A}  B _ \m ^B -
A _ \m B _ \n ^ A + A _ \n B _ \m ^ A \ .
\end{align}
\noindent
Their transformations have the following form

\begin{table}[H]
\centering
\bt{c | cccc}
$ $ &$e ^ {'A} _ {\m \n}$ & $\omega ^ {' AB} _ {\m \n} $ & $A' _ {\m \n}$ & $B ^ {'A} _ {\m \n}$ \\
\hline
$J$ & $ e _ {\m \n} ^ B \Lambda _ B ^ {~A}$ & $\omega ^ {CD} _ {\m \n} \Lambda _ {C} ^ {~A} \Lambda _ {D} ^ {~B}$ & $A _ {\m \n}$ & $ B _ {\m \n} ^ B \Lambda _ B ^ {~A}$ \\
$D$ & $ e ^ {-\a} e _ {\m \n} ^ A $ & $\omega ^ {AB} _ {\m \n}$ & $A _ {\m \n}$ & $ e ^ {\a} B _ {\m \n} ^ A $ \\
$K$ & $ e _ {\m \n} ^ A $ & $\omega _ {\m \n} ^ {AB} + 2 e _ {\m \n} ^ {\l [ A \r.} \a ^ {\l. B \r]} $ & $A _ {\m \n} - 2 e _ {\m \n} ^ C \a _ C$ & $ B _ {\m \n} ^ A + \a ^ {A} _ B e _ {\m \n} ^ B - \omega _ {\m \n} ^ {AB} \a _ B - \a ^ A A _ {\m \n}$
\et
\end{table}

\noindent
We notice that under SCT, the gauge fields mix with the vielbein $e _ \m ^ A$. The origin of this unordinary behavior is the specific form of the commutation relations. According to the rules of the coset construction, the momenta and all the nonlinearly realized generators should form a representation of the group formed by the rest of the generators. Clearly, this condition is broken by the commutation relation between the momenta and conformal boosts~\eqref{conf_cr}.

The transformation properties of the gauge fields would create an obstacle on the way to introducing the covariant derivative for matter fields. However, looking at the transformations of the field strengths,  we see that the expressions simplify considerably once $e _ {\m \n} ^ A = 0$ is imposed. Therefore, as long as pure gravity is concerned, the coset construction produces a sensible result.

The constraint $e _ {\m \n} ^ A$ has the same solution as in the main text; see~\eqref{spin-con3}-\eqref{spin-con5}. The changes appear when one uses also the constraint $e^\n_B\omega ^ {A B}_ {\m \n} = 0$, which can now be solved algebraically in favor of $B_\m^A$. This leads to
\be
B _ \m ^ A e _ {\n A} = (n-2) \l ( \mc R _ {\m \n} - \f {1} {2 (n-1)}g _ {\m \n} \mc R \r ) \ , 
\ee
where $\mc R _ {\m \n} =  \mc R _ {\m \s} ^ {A B} e _ {B} ^ {\s} e _ {A \n}$ and $\mc R=g^{\m\n}\mc R_{\m\n}$ are contractions of the curvature tensor 
\be
\mc R _ {\m \n} ^ {AB}\equiv \bar \omega ^ {AB} _ {\m \n} + \d \omega ^ {AB} _ {\m \n} \ ,
\ee
with $\bar \omega ^ {AB} _ {\m \n} $ and $ \d \omega ^ {AB} _ {\m \n}$ given by~\eqref{barR} and~\eqref{del_omega}.

To obtain the condition for Ricci gauging~\eqref{Ricci_Weyl}, we have to force $B _ \m ^ A$ to vanish. However, it is clear that this constraint is not consistent with SCT. Therefore, in one way or another, we have to dispense of SCT and consider only the gauging of the Poincar\'e group plus dilatations.

\newpage

\section{Irreducible decomposition of torsion \label{irred_decomp}}

We defined the torsion tensor as
\be
\label{tors-append}
T_{\m\n}^A\equiv \p _ \m e _ \n ^ A - \p _ \n e _ \m ^ A - \omega _ {\m B} ^ {A}  e _ \n ^B + \omega _ {\n B} ^ {A}  e _ \m ^B \ , 
\ee
and since it is antisymmetric in $\m$ and $\n$, it has $\frac{n^2(n-1)}{2}$ independent components in an $n$-dimensional spacetime. Under the action of the Lorentz group $SO(1,n-1)$, it can be decomposed into three irreducible quantities:\footnote{In fact, every tensor with the same symmetries as $T_{\m\n}^A$ admits this decomposition.}
\begin{itemize}
\item The vector $\upsilon_\m$
\be
\upsilon_\m=e_{A}^\n T_{\m\n}^A=e^\n_A\l (\p _ \m e _ \n ^ A - \p _ \n e _ \m ^ A + \omega _ {\n B} ^ {A}  e _ \m ^B \r) \ ,
\ee
with $n$ independent components. 

\item The totally antisymmetric ``dual'' tensor 
\be
\a^{\s_1\s_2\cdots \s_{n-3}}=\frac{1}{n \det e}\ep^{\s_1\s_2\cdots \s_{n-3}\m\n\lambda}e_{\lambda A}T_{\m\n}^A \ ,
\ee
with $\frac{n(n-1)(n-2)}{6}$ independent components.
\item The traceless $\frac{n(n^2-4)}{3}$ - component reduced torsion tensor $\tau_{\m\n}^A$
\be
\tau_{\m\n}^A=T_{\m\n}^A-\frac{3}{2(n-1)}\l(\upsilon_\m e_\n^A-\upsilon_\n e^A_\m \r)-\frac{1}{2}e^{\lambda A}\l(T^B_{\lambda\m} e_{\n B}-T^B_{\lambda\n} e_{\m B}\r) \ ,
\ee
which is subject to the following $n+\frac{n(n-1)(n-2)}{6}$ constraints 
\be
\label{constraints-tors}
E^\n_A\tau_{\m\n}^A=0 \ \ \  \text{and} \ \ \ \ep^{\s_1 \s_2 \cdots \s_{n-3}\m\n\lambda} E_{\lambda A}\tau_{\m\n}^A =0 \ .
\ee
\end{itemize}
It is a straightforward exercise to show that~\eqref{tors-append} can be written in terms of the irreducible pieces we presented above as
\be
\begin{aligned}
\label{tors-append2}
T_{\m\n}^A&=\frac{n}{6(n-3)!}\det e~ e^{\lambda A}\ep_{\s_1\s_2\ldots \s_{n-3}\m\n\lambda}\a^{\s_1\s_2\cdots \s_{n-3}}\\
&+\frac{1}{n-1}\l(\upsilon_\m e_\n^A-\upsilon_\n e^A_\m \r)+\frac{2}{3}\tau_{\m\n}^A \ .
\end{aligned}
\ee
Notice that these expressions for  $n=4$ boil down to the ones in~\cite{Diakonov:2011fs}.

\section{Paneitz-Riegert operator}
\label{riegert}

The  Weyl covariant generalization of $\Box ^ 2$ is the Paneitz operator whose form in $n$ dimensions $(n\neq 2)$ was given in~\eqref{pan-rieg}. Using the definition of the Schouten tensor~\eqref{tensor_R}, this operator can be written in a more familiar form as
\be
\begin{aligned}
\mc Q (g) = \nabla ^ 2 &+ \nabla ^ \m  \l[\l(\frac{4} {n-2} R _ {\m \n} - \frac{n^2-4n+8}{2(n-1)(n-2)} g _ {\m \n}R \r ) \nabla ^ \n\r] - \frac{n-4} {4 (n-1)} \nabla ^ 2 R \\
&- \frac{n-4}{(n-2)^2} R _ {\m \n} R ^ {\m \n} + \frac{(n-4)(n^3-4n^2+16n-16)}{16 (n-1)^2(n-2)^2} R ^ 2 \ .
\end{aligned}
\ee
It is interesting to note that for $n=4$, the above expression simplifies considerably
\be
\mc Q (g) \to \nabla ^ 2 +2 \nabla ^ \m  \l[\l(R _ {\m \n} - \frac{1}{3} g _ {\m \n}R \r ) \nabla ^ \n\r] \ ,
\ee
and is also known as the Paneitz-Riegert operator~\cite{Fradkin:1981jc,*Fradkin:1981iu,*Fradkin:1982xc,Riegert:1984kt}.

\bibliographystyle{utphys}
\bibliography{weyl_gauging_coset}{}

\end{document}